\documentstyle[12pt]{article}   
\begin{document}
\title{\bf  Charged particle trajectories in a toroidal magnetic and 
rotation-induced electric field around a black hole}
\author{\bf SUJAN SENGUPTA 
\thanks{Present address : Mehta Research Institute of Mathematics 
and Mathematical Physics, Chhatnag Road, Jhusi, Allahabad~221~506, India.}
\thanks{ E-mail : sujan@mri.ernet.in} 
\\ Indian Institute of Astrophysics, Koramangala,
Bangalore~560~034, India}
\date{}
\maketitle
\vskip1in
{\bf Running Title :} Charged particle dynamics around black holes
\clearpage
\begin{abstract} \normalsize \noindent
Trajectories of charged particle in combined poloidal, toroidal magnetic
field and rotation-induced unipolar electric field superposed in
Schwarzschild background geometry have been investigated extensively in the
context of accreting black holes. The main
purpose of the paper is to obtain a reasonably well insight on the effect of
spacetime curvature to the electromagnetic field surrounding  black holes.
 The coupled equations of motion have been solved numerically
and the results have been compared with that for flat spacetime. It is found
that the toroidal magnetic field dominates  the induced electric field
in determining the motion of charged particles in curved spacetime. 
The combined electromagnetic field repels a charged 
particle from the vicinity of a compact massive object and deconfines the
particle from its orbit.  
In the absence of toroidal magnetic field the particle is trapped
in a closed orbit. The major role of gravitation is to reduce the radius of
gyration significantly while the electric field provides an additional force
perpendicular to the circular orbit. 
Although the effect of inertial frame dragging and the effect of 
magnetospheric plasma
have been neglected, the results provide a reasonably well qualitative picture
of the important role played by gravitation in modifying the electromagnetic
field near accreting black holes and hence the results have potentially important
implications on the dynamics of the fluid and the radiation spectrum associated 
with accreting black holes.
\end{abstract}

\clearpage

{\bf 1. INTRODUCTION}

From the nearest stellar object Sun to the furthest cosmological object quasar,
the presence of the magnetic field effectively governs not only various
physical phenomenon but also the evolution of many astrophysical objects.

The influence of the magnetic field on the dynamical flow of the infalling
plasma around a compact object and on the emergent radiation spectrum have long
been realized in the theoretical study of very high energy astrophysical
objects like pulsars, quasars, Active Galactic Nuclei (AGN) and X-ray binaries.
Unfortunately no convincing theory has been well established so far which can
explain the radiation from these sources consistently and correctly. Though the
definite mechanism has not been understood properly, it is generally believed
that the radiation emission could be due to the plasma processes near compact
objects like neutron stars and black holes. Thus it becomes necessary to
consider plasma processes in intense-gravitational-field backgrounds, as such
compact objects would be massive enough to produce significant spacetime
curvature effects. Therefore one has to consider the dynamics of the fluid in
curved spacetime which is quite an involved problem since one requires to solve
the general relativistic magnetohydrodynamics equations.

It is well known that the best way to understand the structure of any field
is to study the dynamics of test particles in that field. If a particle is
charged then it deviates from its geodesic motion and the study of such
trajectories would reveal informations about the influence of the interacting
field on the spacetime geometry and vice-versa.  The most important in the
astrophysical context is the electromagnetic field. In an electromagnetic field a charged
particle gyrates with a definite radius of gyration and that by and large determines the
radiation spectrum of the object. The concept of radius of gyration in classical mechanics
is modified when one considers curved spacetime$^1$ .
Hence it is very important to
investigate the trajectory of charged particles in curved spacetime with different types of electromagnetic field 
configuration.

 For this purpose one should look for solutions of Einstein-Maxwell equations
which are asymptotically flat and have non-zero dipole magnetic field even in
the absence of rotation. Although these systems of equations are formidable to
solve in general, there are some solutions obtained by perturbation technique
under the assumption that the electromagnetic field is weak compared to the
gravitational field and thus it does not affect the basic geometry. This is
achieved essentially by solving the Maxwell equations on a given curved
spacetime background geometry. This is indeed a valid assumption for even the
most intense magnetic field associated with pulsars carries an energy which is
very small compared to the gravitational potential energy on the surface of a
neutron star of one solar mass. Therefore it is considered that the magnetic
field would not affect the spacetime curvature but the curvature could affect
the magnetic field.

With such a consideration Ginzburg and Ozernoi$^2$ , Petterson$^3$, Bicak
and Dvorak$^4$ have obtained solutions for dipole magnetic field on
Schwarzschild background, whereas using a similar approach Chitre and
Vishveshwara$^5$, Petterson$^6$ have obtained solutions for stationary
electromagnetic fields on Kerr background.

Charged particle dynamics in such electromagnetic fields have been extensively
studied by Prasanna and Varma$^7$  for the Schwarzschild background, by
Prasanna and Vishveshwara$^8$  and by Prasanna and Chakraborty$^9$  for the
Kerr background.

Prasanna and Sengupta$^{10}$  have obtained trajectories of single
charged particle in the presence of toroidal magnetic field superposed on the
Schwarzschild background geometry. In that work it is shown that the toroidal
component of the magnetic field repels the particles away and particularly the
incoming ones get ejected in jet-like trajectories. Hence it has been pointed
out that to produce jet like features from the accreting matter in the high
energy astrophysical scenarios, it is very essential to have the toroidal
component of the magnetic field associated with the central gravitating body.

It is known that almost all of the celestial bodies do have a non-zero angular
momentum and therefore one must consider the important role played by the
angular velocity of the central object in determining not only the spacetime
geometry exterior to it but also the electromagnetic field configurations.
For this purpose Kerr metric has been used in the earlier works and the motion
of the particle in the equatorial plane has been investigated. Since the exterior
region in these studies is considered to be absolutely vacuum, the electric 
field induced by rotation of the object is obtained to be quadrapolar in
nature. However this cannot be a realistic situation as it is well known that
in the case of pulsar magnetosphere and in the accretion disc the region in
which the motion of a charged particle is considered, are plasma filled.
As a consequence Ohm's law should be taken into consideration 
within the entire region while determining
the electric field. Due to the Ohmic induction a rotation-induced 
unipolar electric field
must be present in that region if one neglects the higher multipole by
considering comparatively slow rotation.

In the present paper I have
presented the trajectories of a charged particle under the combined poloidal,
toroidal magnetic field and the rotation-induced unipolar electric field
superposed in Schwarzschild background geometry. This is an extension of the work
by Prasanna and Sengupta$^{10}$. The main purpose of the paper
is to obtain a qualitative picture of the role played by toroidal magnetic field in curved
spacetime in the vicinity  black holes. 

{\bf 2. THE ELECTROMAGNETIC FIELD IN CURVED SPACETIME}

{\bf 2.1. The spacetime metric}

Since we are dealing with the exterior geometry of black holes 
we need the vacuum solution to Einstein's field equations as the
self gravitational field of the surrounding plasma is negligible.

For a rotating compact object one may consider the Kerr metric which has the
following form
\begin{eqnarray}
ds^{2} &=& -\frac{\Delta-a^{2}\sin^{2}\theta}{\rho^{2}}c^{2}dt^{2}-2a^{2}\frac{2mr
\sin^{2}\theta}{\rho^{2}}cdtd\phi+\frac{(r^{2}+a^{2})^{2}-a^{2}\Delta\sin^{2}
\theta}{\rho^{2}}\sin^{2}\theta d\phi^{2}+ \nonumber \\*
& & \frac{\rho^{2}}{\Delta}dr^{2}+\rho^{2}d\theta^{2},
\end{eqnarray}
where 
 $$\Delta = r^{2}-2mr+a^{2}, \; \rho^{2}=r^{2}+a^{2}\cos^{2}\theta, \; a=J/m,$$
$J$ being the total angular momentum and $m=\frac{MG}{c^{2}},$ where $M$ is the
total gravitational mass of the object.

For the case of a black hole the effect due to the rapid rotation
upon the spacetime curvature could be important and the Kerr metric may be
more suitable. Nevertheless if one considers black holes with $\frac{2m}{R}
>> \frac{a}{R}$ one can adopt the Schwarzschild metric, for small $\frac{a}{R}$
the Kerr metric reduces to the Schwarzschild metric and the problem becomes
much easier to tackle with under this approximation. In the present investigation 
we consider sufficiently slow rotation of the object so that the effect of rotation on the background geometry
is insignificant. Hence we consider Schwarzschild background geometry. The investigation with Kerr background geometry 
is under progress. 

The Schwarzschild metric is given by :
\begin{eqnarray}
ds^{2}=-(1-\frac{2m}{r})c^{2}dt^{2}+(1-\frac{2m}{r})^{-1}dr^{2}+r^{2}d\theta^{2}
+r^{2}\sin^{2}\theta d\phi^{2},
\end{eqnarray}
where $m=\frac{MG}{c^{2}}$, $M$ being the total gravitational mass of the
object.

{\bf 2.2. The magnetic field}

The vector potential for the poloidal magnetic field in Schwarzschild background
geometry is given by Petterson$^3$ , Prasanna and Varma$^7$ , Wasserman and
Shapiro$^{11}$  as
\begin{eqnarray}
A_{\phi}=-\frac{3\mu\sin^{2}\theta}{8m^{3}}[r^{2}\ln(1-\frac{2m}{r})+2mr+
2m^{2}].
\end{eqnarray}
Using the definition $F_{ij}=(A_{j,i}-A_{i,j})$ we get the components of the
magnetic field in Schwarzschild background geometry as
\begin{eqnarray}
F_{\phi r}=\frac{3\mu\sin^{2}\theta}{4m^{2}}[\frac{r}{m}\ln(1-\frac{2m}{r})
+(1-\frac{2m}{r})^{-1}+1],
\end{eqnarray}
\begin{eqnarray}
F_{\phi\theta}=\frac{3\mu r^{2}\sin\theta\cos\theta}{4m^{3}}[\ln(1-\frac{2m}{r})
+\frac{2m}{r}(1+\frac{m}{r})],
\end{eqnarray}
$\mu$ being the dipole moment associated with the central source.

In magnetized accretion disks around a compact object the surrounding plasma
is supposed to be tied to the magnetic field and its inertia would cause
the magnetic field lines to be bent backward creating a toroidal component.
The toroidal field component in Schwarzschild background geometry 
is given as$^{10}$ 
\begin{eqnarray}
F_{r\theta}=\frac{rB_{\circ}(r_{\circ}-2m)}{r_{\circ}(r-2m)\sin\theta} \;,
\end{eqnarray}
$r_{\circ}$ and $B_{\circ}$ being constants.

{\bf 2.3. The induced electric field}

We consider the electromagnetic field to be stationary. For this we require
the magnetic dipole axis to be aligned with the rotational axis of the central
object. In the present study we ignore any effects due to the surrounding
plasma. Since the surrounding medium exterior to the object is considered to
be plasma filled, there will be a rotation-induced unipolar electric field
such that ${\bf E.B}=0$ everywhere. The components of the induced electric
field considered  here 
 are obtained by using
the generalized Ohm's law (assuming infinite conductivity of the medium)
$$ F^{\beta}_{\delta}u^{\delta}=0,$$
where $u^{\delta}=(u^{t},0,0,u^{\phi})$ are the components of the four velocity
vector of the corotating plasma. 

In Schwarzschild background geometry the components of the induced unipolar
electric field are given as$^{12}$ 
\begin{eqnarray}
F_{rt}=\frac{3\mu\omega\sin^{2}\theta}{4m^{2}c}[\frac{r}{m}\ln(1-\frac{2m}{r})
+(1-\frac{2m}{r})^{-1}+1],
\end{eqnarray}
\begin{eqnarray}
F_{\theta t}=\frac{3\mu\omega r^{2}}{4m^{3}c}\sin\theta\cos\theta[\ln(1-
\frac{2m}{r})+\frac{2m}{r}(1+\frac{m}{r})],
\end{eqnarray}
where $\omega$ is the angular velocity of the central object as measured by
an observer at infinity.

From equations (7) and (8) we obtain the  potential for the electric
field as
\begin{eqnarray}
A_{t}=\frac{3\mu\omega r^{2}\sin^{2}\theta}{8m^{3}c}[\ln(1-\frac{2m}{r})+
\frac{2m}{r}(1+\frac{m}{r})].
\end{eqnarray}

The electric field arises due to the corotation of the plasma with the object
and is contributed by the poloidal magnetic field. The toroidal magnetic field
has no contribution to the electric field as can be seen from the generalized
Ohm's law.

Therefore in the present discussion we are considering a region with combined
poloidal and toroidal magnetic field and rotation-induced unipolar electric
field superposed on Schwarzschild background geometry. 

{\bf 3. THE EQUATIONS OF MOTION OF A CHARGED PARTICLE}

{\bf 3.1. In Schwarzschild background geometry}

For the motion of a charged particle of charge $e$ and rest mass $m_{\circ}$
there are two constants of motion, the canonical angular momentum $l$ and
the energy $E$ as given by
\begin{eqnarray}
v_{\phi}+eA_{\phi}=l,
\end{eqnarray}
\begin{eqnarray}
v_{t}+eA_{t}=-E,
\end{eqnarray}
wherein all quantities are normalized with respect to the particle rest mass
$m_{\circ}$. The velocity four vector of the charge particle, $v^{i}$ is
timelike, which for the metric of signature +2 is expressed as
$$g_{ij}v^{i}v^{j}=-1.$$

From equation (10) we get
\begin{eqnarray}
v^{\phi}=\frac{d\phi}{ds}=\frac{l}{r^{2}\sin^{2}\theta}+\frac{3e\mu}{8m_{\circ}
c^{2}m^{3}}[\ln(1-\frac{2m}{r})+\frac{2m}{r}(1+\frac{m}{r})].
\end{eqnarray}

From equation (11) we get
\begin{eqnarray}
v^{t}=\frac{cdt}{ds}=(1-\frac{2m}{r})^{-1}\{E+\frac{3e\mu\omega 
r^{2}\sin^{2}\theta}{8m^{3}c^{3}m_{\circ}}[\ln(1-\frac{2m}{r})
+\frac{2m}{r}(1+\frac{m}{r})]\}.
\end{eqnarray}

From the covarient Lorentz equations
$$v^{i}_{;j}v^{j}=eF_{j}^{i}v^{j},$$
where $v^{i}=\frac{dx^{i}}{ds}$ is the four velocity of the particle and the
semicolon denotes the covariant derivative with respect to the spacetime
metric, we obtain
\begin{eqnarray}
\frac{d^{2}r}{ds^{2}} &=&
\frac{m}{r^{2}}(1-\frac{2m}{r})^{-1}(\frac{dr}{ds})^{2}
+r(1-\frac{2m}{r})\{(\frac{d\theta}{ds})^{2}+\sin^{2}\theta(\frac{d\phi}{ds}
)^{2}\}- \nonumber \\*
& & \frac{m}{r^{2}}(1-\frac{2m}{r})(\frac{cdt}{ds})^{2}+\frac{e}{m_{\circ}c^2}
(1-\frac{2m}{r})\{\frac{rB_{\circ}(r_{\circ}-2m)}{r_{\circ}(r-2m)\sin\theta}
\frac{d\theta}{ds}- \nonumber \\*
& & \frac{3\mu\sin^{2}\theta}{2m^{2}}[(1-\frac{m}{r})+(\frac{r}{2m}-1)\ln(1-
\frac{2m}{r})](1-\frac{2m}{r})^{-1}(\frac{d\phi}{ds}-\omega\frac{dt}{ds})\},
\end{eqnarray}
\begin{eqnarray}
\frac{d^{2}\theta}{ds^{2}} &=& -\frac{2}{r}\frac{dr}{ds}\frac{d\theta}{ds}+
\sin\theta\cos\theta(\frac{d\phi}{ds})^{2}-\frac{e}{m_{\circ}c^{2}r^{2}}
[\frac{rB_{\circ}(r_{\circ}-2m)}{r_{\circ}(r-2m)\sin\theta}\frac{dr}{ds}
- \nonumber \\*
& & \frac{3\mu\sin\theta\cos\theta}{4m^3}\{r^{2}
\ln(1-\frac{2m}{r})+2m(r+m)\}(\omega\frac{dt}{ds}-\frac{d\phi}{ds})].
\end{eqnarray}

{\bf 3.2. In flat spacetime}

In order to investigate the effects of spacetime curvature we need to determine
the trajectories of the charge particle in flat spacetime as well. These will 
provide the motion of a particle in the Newtonian mechanics under various
electromagnetic conditions. However in the present discussion we neglect the
toroidal magnetic field component for flat spacetime since the poloidal
magnetic and the induced electric fields are sufficient for a comparative
study of the particle motion in curved and flat spacetimes. 

Expanding in Taylor series and neglecting the higher order terms containing
$m$ we obtain from equations (12) to (15) the equations of motion of a charge
particle in electromagnetic field without gravitation as
\begin{eqnarray}
\frac{d^{2}r}{ds^{2}}=r\{(\frac{d\theta}{ds})^{2}+\sin^{2}\theta(\frac{d\phi}{d
s})^{2}\}-\frac{e\mu\sin^{2}\theta}{m_{\circ}c^{2}r^{2}}(\frac{d\phi}{ds}-
\omega\frac{dt}{ds}) \; ,
\end{eqnarray}
\begin{eqnarray}
\frac{d^{2}\theta}{ds^{2}}=-\frac{2}{r}\frac{dr}{ds}\frac{d\theta}{ds}+\sin
\theta\cos\theta(\frac{d\phi}{ds})^{2}-\frac{2e\mu\sin\theta\cos\theta}{m_{
\circ}c^{2}r^{3}}(\omega\frac{dt}{ds}-\frac{d\phi}{ds}) \; ,
\end{eqnarray}
\begin{eqnarray}
\frac{d\phi}{ds}=\frac{l}{r^{2}\sin^{2}\theta}-\frac{e\mu}{m_{\circ}c^{2}r^3} \; ,
\end{eqnarray}
\begin{eqnarray}
\frac{cdt}{ds}=E-\frac{e\mu\omega\sin^{2}\theta}{m_{\circ}c^{3}r} \; .
\end{eqnarray}

{\bf 4. NUMERICAL SOLUTIONS}

As the orbit equations involve transcendental functions it is analytically
impossible to make any general analysis. Hence we restore to numerical
integration and get the qualitative picture of the nature of trajectories
for different values of the physical parameters appearing in the equations. For
convenience of calculations we shall consider the equations in dimensionless
form by introducing the dimensionless quantities
$$\rho=\frac{r}{m}, \; \sigma=\frac{s}{m}, \; L=\frac{l}{m}, \; \tau=\frac{ct}{m},
\; W=\frac{m\omega}{c}, \; \lambda_{T}=\frac{eB_{\circ}}{m_{\circ}c^2}, \;
\lambda_{P}=\frac{e\mu}{m_{\circ}c^{2}m^{2}} \; .$$
With this definition the equations of motion for curved spacetime read as
\begin{eqnarray}
\frac{d^{2}\rho}{d\sigma^{2}} &=& \frac{1}{\rho^{2}}(1-\frac{2}{\rho})^{-1}
(\frac{d\rho}{d\sigma})^{2}+\rho(1-\frac{2}{\rho})\{(\frac{d\theta}{d\sigma}
)^{2}+\sin^{2}\theta(\frac{d\phi}{d\sigma})^{2}\} \nonumber \\*
& & -\frac{1}{\rho^{2}}(1-\frac{2}{\rho})(\frac{d\tau}{d\sigma})^{2}+
\frac{\Lambda\lambda_{P}(1-\frac{2}{\rho_{\circ}})}{\sin\theta}\frac{d\theta}{d
\sigma}- \nonumber \\*
& & \frac{3\lambda_{P}}{2}\sin^{2}\theta\{(1-\frac{1}{\rho})+(\frac{\rho}{2}-1)
\ln(1-\frac{2}{\rho})\}(\frac{d\phi}{d\sigma}-W\frac{d\tau}{d\sigma}) \; ,
\end{eqnarray}
\begin{eqnarray}
\frac{d^{2}\theta}{d\sigma^{2}} &=& -\frac{2}{\rho}\frac{d\rho}{d\sigma}
\frac{d\theta}{d\sigma}+\sin\theta\cos\theta(\frac{d\phi}{d\sigma})^{2}-
\frac{\Lambda\lambda_{P}(1-\frac{2}{\rho_{\circ}})}{\rho^{2}(1-\frac{2}{\rho})
\sin\theta}\frac{d\rho}{d\sigma} \nonumber \\*
& & +\frac{3\lambda_{P}}{4}\sin\theta\cos\theta\{\ln(1-\frac{2}{\rho})+
\frac{2}{\rho}(1+\frac{1}{\rho})\}(W\frac{d\tau}{d\sigma}-\frac{d\phi}{d\sigma}) \; ,
\end{eqnarray}
\begin{eqnarray}
\frac{d\phi}{d\sigma}=\frac{L}{\rho^{2}\sin^{2}\theta}+\frac{3\lambda_{P}}{8}
[\ln(1-\frac{2}{\rho})+\frac{2}{\rho}(1+\frac{1}{\rho})] \; ,
\end{eqnarray}
\begin{eqnarray}
\frac{d\tau}{d\sigma}=(1-\frac{2}{\rho})^{-1}\{E+\frac{3\lambda_{P}W\rho^2}{8}
\sin^{2}\theta[\ln(1-\frac{2}{\rho})+\frac{2}{\rho}(1+\frac{1}{\rho})]\} \; ,
\end{eqnarray}
where $\Lambda=\frac{\lambda_{T}}{\lambda_{P}}$.

Similarly, the equations of motion for flat spacetime read as (without the
toroidal field components)
\begin{eqnarray}
\frac{d^{2}\rho}{d\sigma^{2}}=\rho\{(\frac{d\theta}{d\sigma})^{2}+\sin^{2}\theta
(\frac{d\phi}{d\sigma})^{2}\}-\frac{\lambda_{P}}{\rho^{2}}\sin^{2}\theta(
\frac{d\phi}{d\sigma}-W\frac{d\tau}{d\sigma}) \; ,
\end{eqnarray}
\begin{eqnarray}
\frac{d^{2}\theta}{d\sigma^{2}}=-\frac{2}{\rho}\frac{d\rho}{d\sigma}
\frac{d\theta}{d\sigma}+\sin\theta\cos\theta(\frac{d\phi}{d\sigma})^{2}-
\frac{2\lambda_{P}}{\rho^{3}}\sin\theta\cos\theta(W\frac{d\tau}{d\sigma}
-\frac{d\phi}{d\sigma}) \; ,
\end{eqnarray}
\begin{eqnarray}
\frac{d\phi}{d\sigma}=\frac{L}{\rho^{2}\sin^{2}\theta}-
\frac{\lambda_{P}}{\rho^3} \; ,
\end{eqnarray}
\begin{eqnarray}
\frac{d\tau}{d\sigma}=E-\frac{\lambda_{P}W}{\rho}\sin^{2}\theta.
\end{eqnarray}

In order to solve the equations of motion we need to pescribe the initial
conditions.

We assume initially $\rho=\rho_{\circ}, \; \theta=\theta_{\circ}, \; 
\phi=\phi_{\circ} \; and \; \tau=\tau_{\circ}.$
Also at $\rho_{\circ}, \; \frac{d\theta}{d\sigma}=
(\frac{d\theta}{d\sigma})_{\circ}, \;
\frac{d\phi}{d\sigma}=(\frac{d\phi}{d\sigma})_{\circ}$.

Throughout our discussion we take $(\frac{d\phi}{d\sigma})_{\circ}=0$ and
$\tau_{\circ}=0$.
Therefore, from equation (22) we get
\begin{eqnarray}
L=-\frac{3\lambda_{P}\rho^{2}_{\circ}}{8}\sin^{2}(\theta_{\circ})[\ln(1-
\frac{2}{\rho_{\circ}})+\frac{2}{\rho_{\circ}}(1+\frac{1}{\rho_{\circ}})].
\end{eqnarray}

From equation (23) we get
\begin{eqnarray}
(\frac{d\tau}{d\sigma})_{\circ}=(1-\frac{2}{\rho_{\circ}})^{-1}\{E+
\frac{3\lambda_{P}W\rho_{\circ}^{2}}{8}\sin^{2}(\theta_{\circ})[\ln(1-\frac{2}{\rho_{\circ}})
+\frac{2}{\rho_{\circ}}(1+\frac{1}{\rho_{\circ}})]\}.
\end{eqnarray}

From the spacetime metric (2) we get
\begin{eqnarray}
(\frac{d\rho}{d\sigma})_{\circ}=\pm[(1-\frac{2}{\rho_{\circ}})\{(1-
\frac{2}{\rho_{\circ}})(\frac{d\tau}{d\sigma})^{2}_{\circ}-\rho^{2}_{\circ}
(\frac{d\theta}{d\sigma})^{2}_{\circ}+1\}]^{1/2}.
\end{eqnarray}

Equations (28) - (30) provide the initial conditions for the motion of the
particle in curved spacetime.

Similarly, from equation (26) we obtain
\begin{eqnarray}
L=\frac{\lambda_{P}}{\rho_{\circ}}\sin^{2}(\theta_{\circ}).
\end{eqnarray}
From equation (27) we get
\begin{eqnarray}
(\frac{d\tau}{d\sigma})_{\circ}=E-\frac{\lambda_{P}W}{\rho_{\circ}}\sin^{2}
(\theta_{\circ}).
\end{eqnarray}

From the Minkowskian metric
$$ds^{2}=-c^{2}dt^{2}+dr^{2}+r^{2}d\theta^{2}+r^{2}\sin^{2}\theta d\phi^{2},$$
we get
\begin{eqnarray}
(\frac{d\rho}{d\sigma})_{\circ}=\pm[1+(\frac{d\tau}{d\sigma})_{\circ}^{2}-
\rho^{2}_{\circ}(\frac{d\theta}{d\sigma})^{2}_{\circ}]^{1/2}.
\end{eqnarray}

Equations (31) - (33) give the initial conditions for the motion of the
particle in flat spacetime.

The equations are numerically solved by using the 4th order adaptive step size
Runga-Kutta method. The integrations are performed between finite limits with
an accuracy of $10^{-7} - 10^{-8}$, while changing the step size appropriately
through self inspection depending upon the initial conditions. The trajectories
are obtained in spherical polar co-ordinates which are then converted into
Cartesian co-ordinates by using the relations
$$x=r\sin\theta\cos\phi, \; y=r\sin\theta\sin\phi, \; z=r\cos\theta.$$
The results are presented graphically. 

{\bf 5. RESULTS AND DISCUSSIONS}

As obtained by Prasanna and Varma$^7$ , when the magnetic field is large, the
motion of a charged particle looks similar to that in the flat geometry wherein
the particle gyrates in a given tube of lines reflecting between two mirror
points located symmetrically with respect to the equatorial plane (see Figure
1a). On the other hand, if the magnetic field is lower, then the particle
oscillations up and down the equatorial plane damp continuosly as r decreases
and eventually the particle gets sucked in by the central object. The inclusion
of the induced unipolar electric field changes the nature of the orbit by
introducing a ${\bf E\times B}/B^2$ drift. The particle still remains trapped
in a mirror configuration reflecting between two points on either side of the
equatorial plane (see Figure 1b) just like the case without the electric field.
Figure 1c shows the trajectory of the particle in a combined poloidal and
unipolar electric field superposed in flat spacetime. A comparison between
Figure 1b and 1c reveals that the gravitational field mainly contributes in
decreasing the radius of gyration of the particle significantly i.e., in curved
spacetime the particle gyrates in much tighter orbit than it does in flat
spacetime.

From equation (12) to equation (15) we notice that unlike the case of pure
poloidal field, a particle on the plane $\theta = \pi/2$ having zero initial
velocity in the meridonial direction is not confined to the equatorial plane as
the acceleration along the $\theta$ direction is still non-zero in the presence
of toroidal magnetic field. The inclusion of rotational motion of the object
modifies the initial conditions. From the initial conditions we notice that
the initial velocity of the particle is well defined even if $E=0$ since the
particle acquires its initial velocity due to its interaction with the electric
field. The situation is similar to the case for the
Kerr background geometry$^8$. But here
we have considered a rigid rotation neglecting the dragging of inertial frame
effect. Since it is difficult to obtain the trajectory of a charged particle
in the off equatorial plane with the Kerr geometry due to the presence of 
non-zero off  diagonal terms in the metric, the particle trajectory has not
been investigated in that situation although the situation is more realistic
in that case for astrophysical interest. The present study can very well provide
a nearly similar qualitative picture with two differences i.e., the absence of
the dragging of inertial frame effect and the rotation-induced quadrapolar 
electric field. As mentioned in the introduction, if we consider comparatively
slow rotation, the inertial frame dragging becomes negligible and the
consideration of rotation-induced unipolar electric field is much more
realistic in the astrophysical situation than that of qudrapolar electric
field.

Prasanna and Sengupta$^{10}$  investigated the trajectories of a charged
particle in combined poloidal and toroidal  magnetic field superposed in
Schwarzschild background geometry. The effect of rotation and hence the
electric field has not been considered in that study. In that work it is
found that when the toroidal field is extremely small compared to the poloidal
field the particle trajectory appears somewhat similar to that in the case
of pure poloidal field till the particle reaches a distance two Schwarzschild
radii away from the central object where gravity seems to dominate and pull the
particle in. As the magnitude of the toroidal field is increased, the nature of
the trajectory alters and when the magnitude of the toroidal field 
becomes significant compared to that of the poloidal field
the particle gets deconfined even very near to the central
gravitating object and gets pushed away to infinity. The most interesting
feature is that if the initial velocity of the particle is towards the central
object then in the presence of the toroidal field the particle comes very near
to the object and then gets ejected away to infinity.

As the earlier investigation has not included the electric field induced by
rotation one may expect significant changes in the trajectories in more
realistic case. But here we show that the above situation remains the same even
if one includes the electric field. Figure 2 and Figure 3a show the trajectory
of a charged particle in combined toroidal, poloidal magnetic field and induced
unipolar electric field superposed in Schwarzschild background geometry under
different initial conditions. In these situations the particle starting at any
given position and having its radial velocity inwards comes towards the central
object slowly and as it approaches its minimum value of r (depending on
$r_{\circ}$ ) and of $\theta$ (depending upon the angular momentum) the
particle gets bounced off in a straight line trajectory. This phenomenon could
be a potential mechanism for the acceleration of high energy cosmic ray
particles. In case the particle's radial velocity is directly away from the
center, the particle spirals around the central
body as depicted in Figure 3a. It is very interesting to note that near the object
 the nature of the particle trajectory remains the same  (as shown in Figure 2)
whether one includes the electric field or not.
However, at a very large distance from the central object the presence of
the electric field becomes visible as the two trajectories differ significantly.
 In the case for the closed orbit the trajectory remains almost the same 
 with and without the presence of the electric field as observed from 
Figure 3a and Figure 3b. Thus one is lead to a 
very interesting conclusion that the presence of toroidal magnetic field 
totally dominates the role of the electric field in determining the motion
 of a charged particle in the presence of a strong gravitational field.

{\bf 6. CONCLUSIONS}

In order to understand the effect of curved spacetime on the electromagnetic
field surrounding compact objects the trajectories of charged particles
have been investigated in the context of accretion onto black holes.
For this purpose the trajectories of a charged
particle in combined poloidal, toroidal magnetic field and rotation-induced
 unipolar electric field superposed in Schwarzschild background
geometry have been obtained. It is found that the toroidal component of
the magnetic field dominates the role of unipolar electric field in determining
the motion of a charged particle in the presence of strong gravitational field.
In the presence of a non-zero toroidal magnetic field an incoming particle
approaches very near to the central object and then gets ejected away to
infinity. This phenomenon could be a plausible mechanism
for the production of high energy cosmic ray. A rigorous magnetohydrodynamical
investigation including the toroidal component of the magnetic field 
superposed in curved spacetime could give rise to the production of jet like
phenomenon observed in quasars and AGN. In the absence of toroidal field the particle
is trapped into a closed orbit in the same way it behaves in the presence of
only poloidal magnetic field except a regular oscillation perpendicular to
the orbital motion is produced by the presence of the electric field.

{\bf ACKNOWLEDGEMENTS}

I express my gratitude to A. R. Prasanna for introducing me into this
subject. Thanks are due to C. V. Vishveshwara and B. R. Iyer for their
help and encouragement. 

\clearpage
{\bf REFERENCES}
\begin{tabbing}
1. M. A. Abramowicz, J. C. Miller, Z. Stuchilk, (1993). Phys. Rev. {\bf D47}, 1440.\\
2. V. L. Ginzburg and I. M. Ozernoi, (1965). Sov. Phys. JETP {\bf 20}, 689. \\
3. J. A. Petterson, (1975) Phys. Rev. {\bf D10}, 3166. \\
4. J. Bicak and L. Dvorak, (1977). Czech. J. Phys. {\bf B27}, 127. \\
5. D. M. Chitre and C. V. Vishveshwara, (1975). Phys. Rev. {\bf D12}, 1538. \\
6.  J. A. Petterson, (1975). Phys. Rev. {\bf D12}, 2218. \\
7. A. R. Prasanna and R. K. Varma, (1977). Pramana, {\bf 8}, 229. \\
8. A. R. Prasanna and C. V. Vishveshwara, (1978). Pramana, {\bf 11}, 359. \\
9. A. R. Prasanna and D. K. Chakraborty, (1980). Pramana, {\bf 14}, 113. \\
10. A. R. Prasanna and S. Sengupta, (1994). Phys. Letts. A {\bf 193}, 25. \\
11. I. Wasserman and S. L. Shapiro, (1983). Astrophys. J., {\bf 265}, 1036. \\
12. S. Sengupta, (1995). Astrophys. J., {\bf 449}, 224. 
\end{tabbing}
\clearpage
\begin{center}
{\bf Figure Captions}
\end{center}
{\bf Fig. 1a. --} Projection of the trajectory of a positively charged particle
(in purely poloidal magnetic field superposed on curved spacetime) in XY plane.
The various parameters are $E=2.0$, $L=91.77$, $\lambda_{P}=
80$, $\rho_{\circ}=2.5$, $(\frac{d\rho}{d\sigma})_{\circ}=-1.92$, 
$\theta_{\circ}=\pi/2$, $(\frac{d\theta}{d\sigma})_{\circ}=0.3$, $\phi_{\circ}
=0^{\circ}$. In all figures a positive value of
$(\frac{d\rho}{d\sigma})_{\circ}$ implies that the initial motion of the
particle to be away from the center whereas a negative value to be towards the
center.

{\bf Fig. 1b. --} Projection of the trajectory of a positively charged particle
in poloidal magnetic and unipolar electric field superposed in curved
spacetime. The parameters are the same as used for Figure 1a except
$(\frac{d\rho}{d\sigma})_{\circ}=-1.122$ and $W=0.01$.

{\bf Fig. 1c. --} Same as Figure 1b but without gravitational field. The
parameters are the same except $(\frac{d\rho}{d\sigma})_{\circ}=-1.8$ and
$L=32.0$.

{\bf Fig. 2. --} Projection of the trajectory of a charged particle on XY,
XZ and YZ planes. The solid line represents the trajectory in the
combined poloidal, toroidal magnetic and unipolar electric field superposed
in curved spacetime while the dashed line represents that without the 
electric field.  
The parameters are $E=5.0$, $L=17.9$, $\lambda_{P}=80.0$, 
$\lambda_{T}=4.0$, $\rho_{\circ}=6.0$, 
$\theta_{\circ}=\pi/2$, $(\frac{d\theta}{d\sigma})_{\circ}=0.0$, $W=0.01$ and 
$\phi_{\circ}=0.0$. With the electric field $(\frac{d\rho}{d\sigma})_{\circ}
=-4.89, $ without the electric field $(\frac{d\rho}{d\sigma})_{\circ}= -4.93.$

{\bf Fig. 3a. --} Projection of the trajectory of a charged particle in XY
 plane in the combined poloidal, toroidal magnetic and unipolar electric
field superposed in curved spacetime. The parameters are the same as given for
Figure 2 but $\rho_{\circ}=3.5$,
$(\frac{d\rho}{d\sigma})_{\circ}=+4.63$, $L=41.382$,
and $\lambda_{T}=40$.

{\bf Fig. 3b. --} Projection of the trajectory of a charged particle in XY
 plane in the combined poloidal, toroidal magnetic but without unipolar
 electric  field. The parameters are the same as given for
Figure 3a except  $(\frac{d\rho}{d\sigma})_{\circ}=+4.96$.
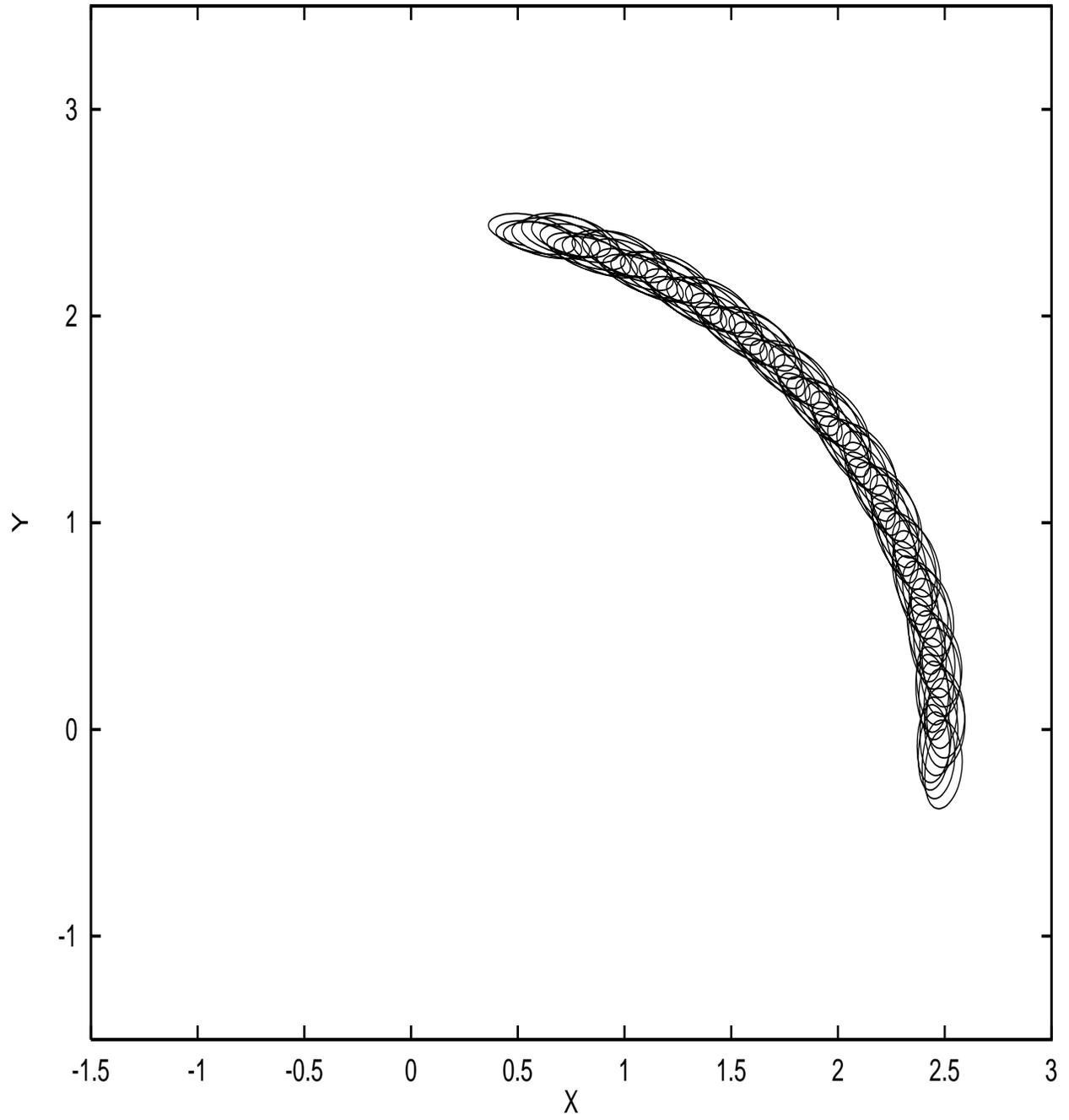
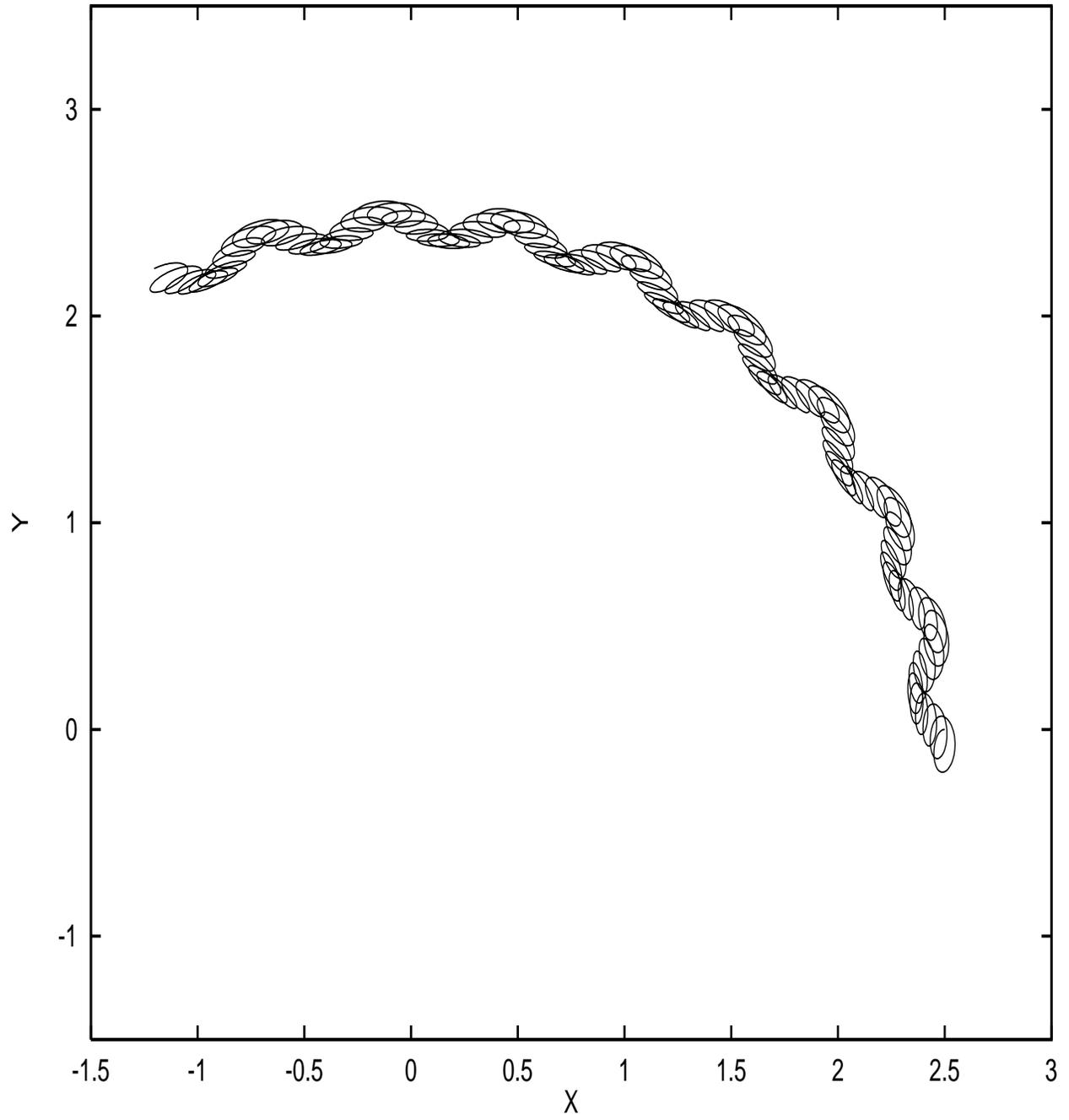
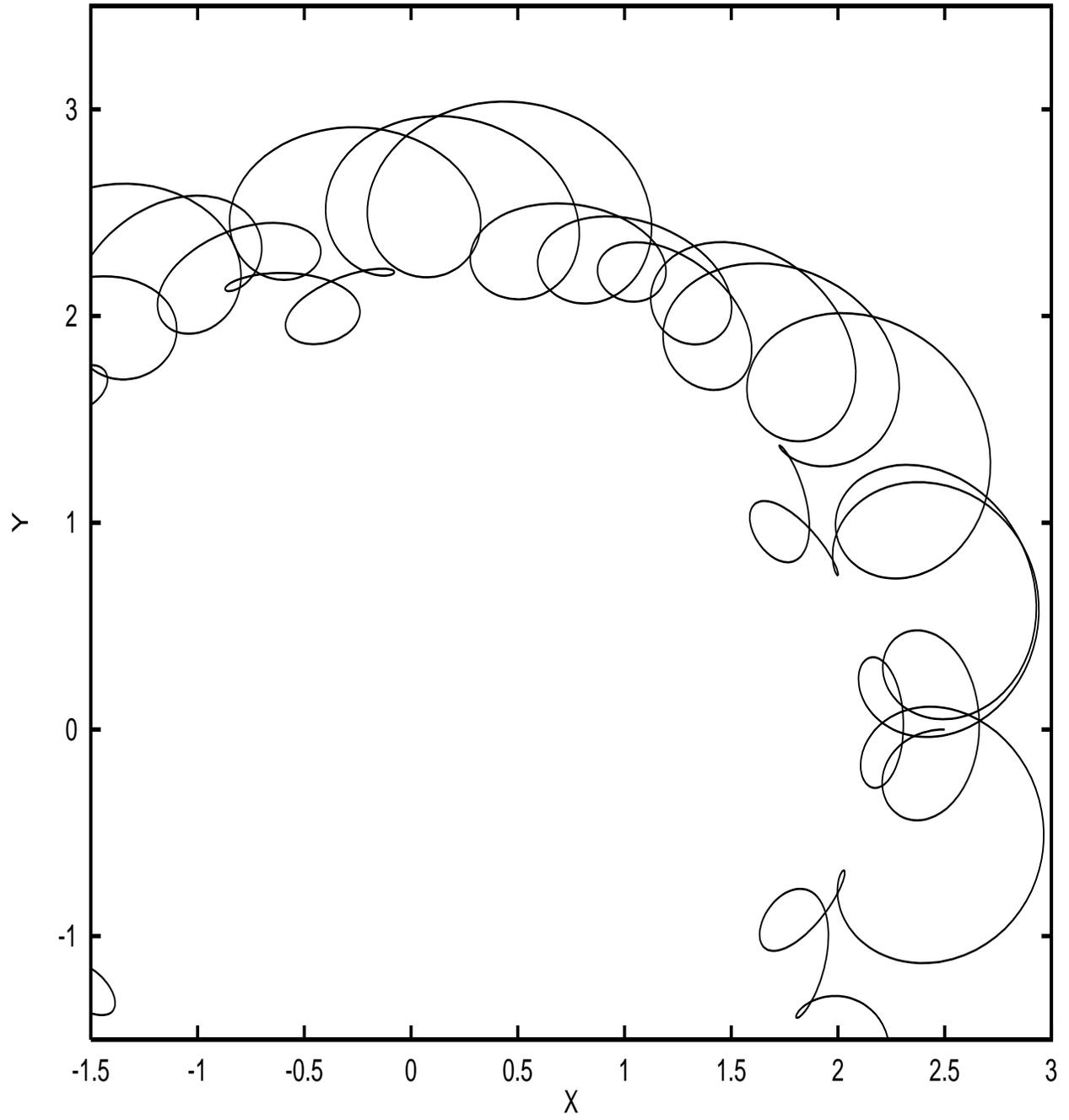
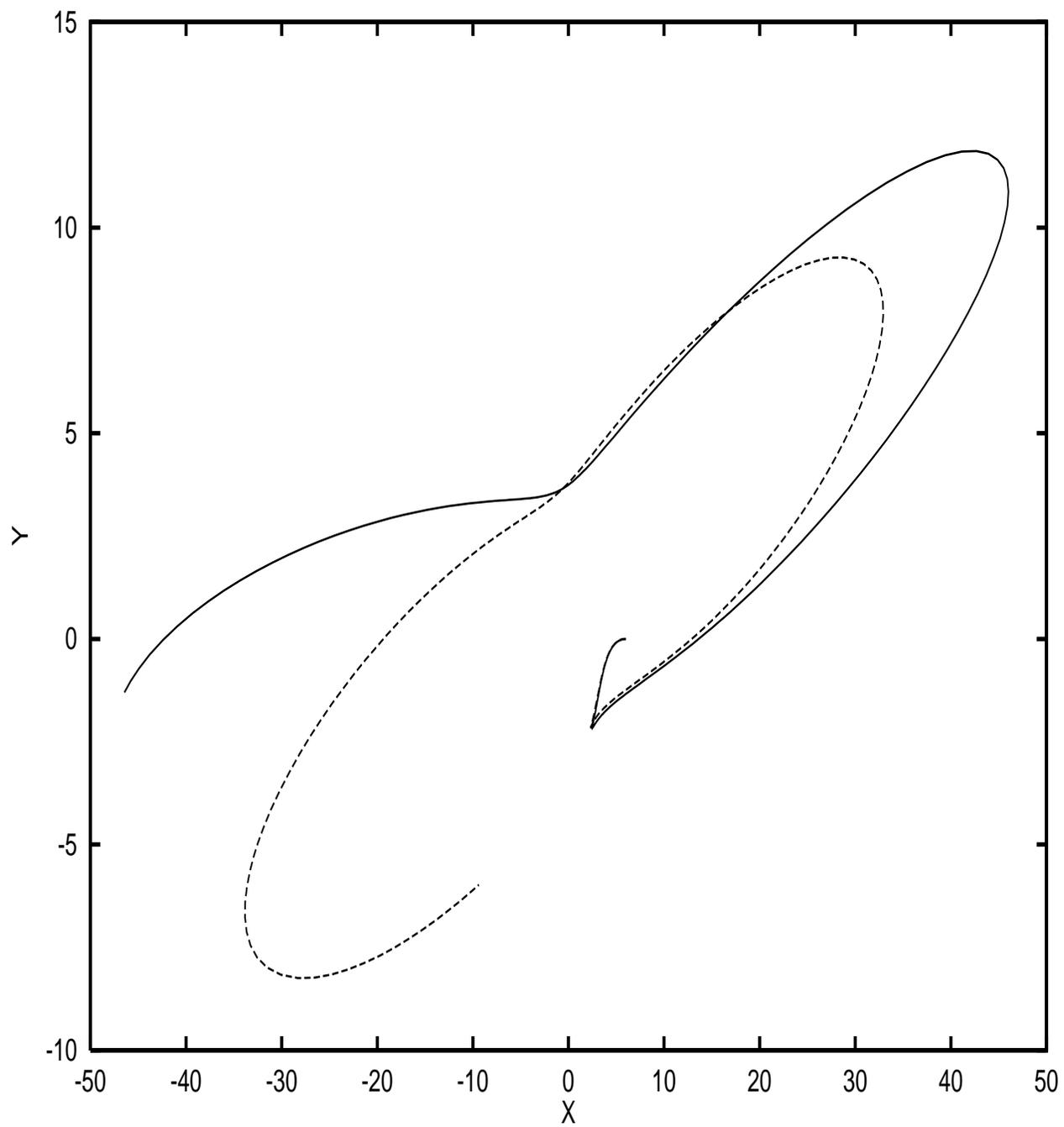
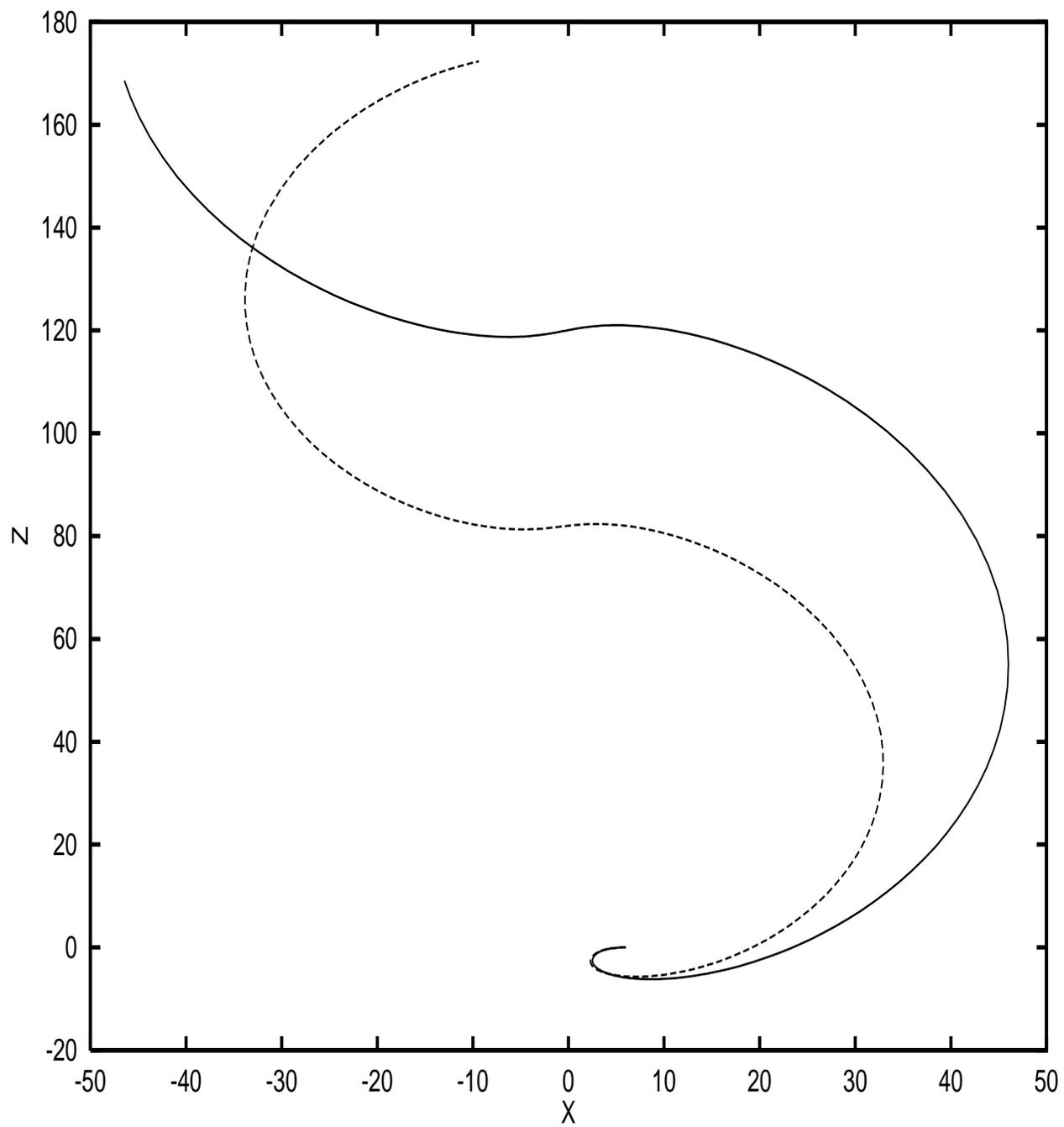
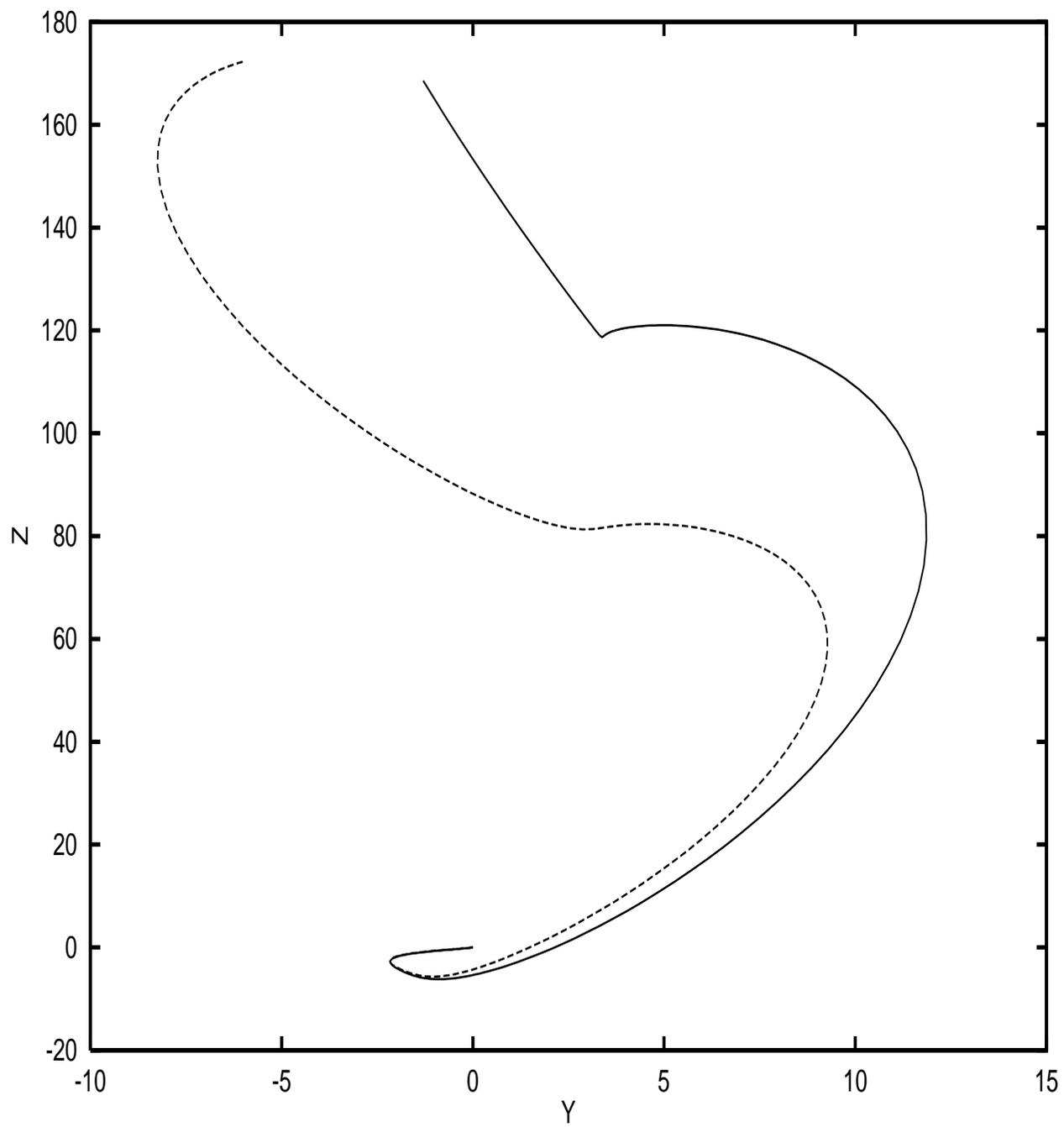
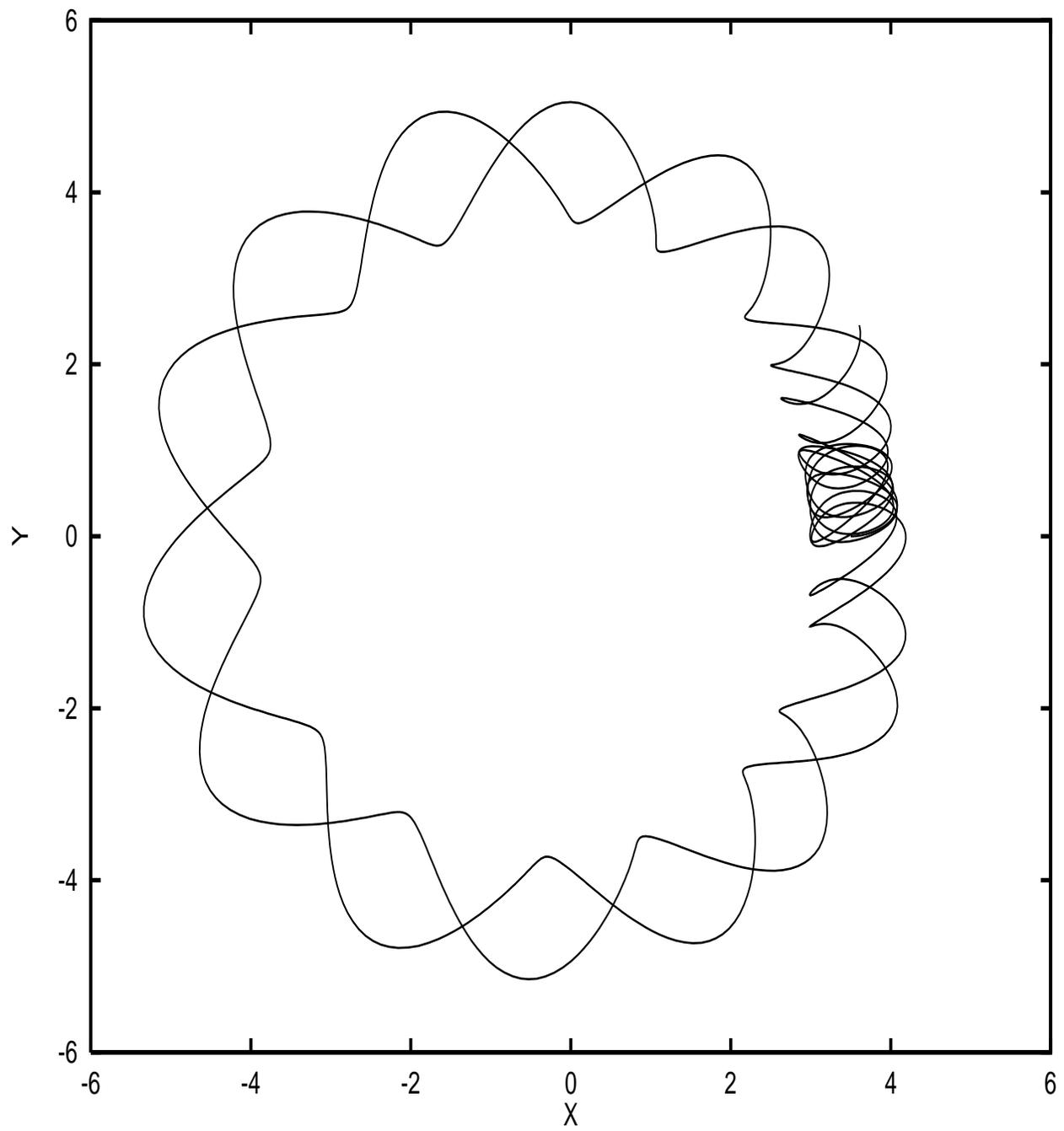
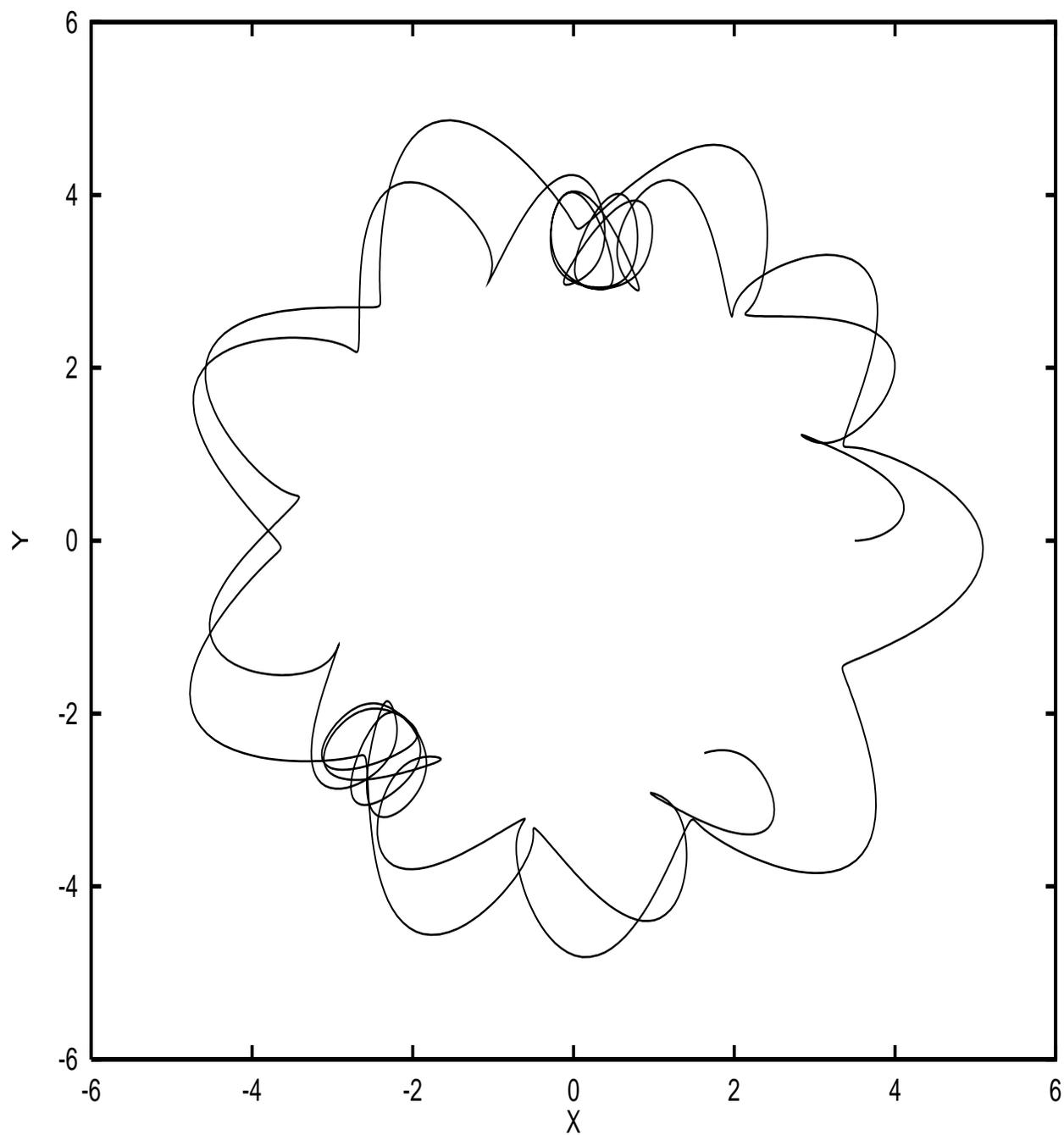
\end{document}